\documentclass[preprint,12pt]{elsarticle}

\usepackage{amssymb}

\usepackage{tabularx}
\usepackage{multirow}
\usepackage{url}
\usepackage[ansinew]{inputenc}
\usepackage{amsmath}
\usepackage{amssymb}
\usepackage[
]{graphicx}
\usepackage{color}
\usepackage{subfig}

\newcommand{\amp}{A}
\newcommand{\eigf}{f}

\DeclareMathOperator{\dg}{d}

\renewcommand{\Re}[1]{\operatorname{Re}\left\{ #1 \right\}}

\graphicspath{{pic/}}

\journal{International Journal of Aeroacoustics}

\begin{document}

\begin{frontmatter}


\title{A robust extension to the triple plane pressure mode matching method 
	by filtering convective perturbations}


\author[dlr]{Attila Wohlbrandt\corref{cor1}}
\ead{attila.wohlbrandt@dlr.de}
\author[dlr]{Christian Weckm\"uller}
\ead{christian.weckmueller@mail.de}
\author[dlr]{S\'ebastien Gu\'erin}
\ead{sebastien.guerin@dlr.de}

\cortext[cor1]{Corresponding author}
\address[dlr]{German Aerospace Center, Institute of Propulsion Technology, Engine Acoustics Dept., 
M\"uller-Breslau-Strasse 8, 10623 Berlin, Germany}

\begin{abstract}

Time-periodic CFD simulations are widely used to investigate turbomachinery components. 
The triple-plane pressure mode matching method (TPP) developed by Ovenden and
Rienstra extracts the acoustic part in such simulations.
Experience shows that this method is subject to significant errors when the
amplitude of pseudo-sound is high compared to sound. 
Pseudo-sound are unsteady pressure fluctuations with a convective character.
The presented extension to the TPP improves the splitting between acoustics and the rest of the unsteady flow field. The method is simple:  
i) the acoustic eigenmodes are analytically determined for a uniform mean flow as in the original TPP; 
ii) the suggested model for convective pressure perturbations uses the convective wavenumber as axial wavenumber and the same orthogonal radial shape functions as for the acoustic modes. 
The reliability is demonstrated on the simulation data of a low-pressure fan. 
As acoustic and convective perturbations are separated, the
accuracy of the results increases close to sources, allowing a
reduction of the computational costs by shortening the simulation domain. 
The extended method is as robust as the original one--giving the same results for the acoustic modes in absence of convective perturbations.

\end{abstract}


\end{frontmatter}



\section{\MakeUppercase{Introduction}}
\label{sec:introduction}
CFD simulations sustain the development and design of new turbomachinery components by providing valuable and detailed information. 
The prediction of tonal fan noise using time-accurate RANS (URANS) or Harmonic
Balance is by now an established method.
In many cases the acoustic levels are extracted upstream and downstream of the source inside the duct without a costly propagation into the far field. 


An overview and discussion of existing methods for the acoustic post-processing applied to rotor--stator interaction can be found in Giacch\'e et al.~\cite{Giacche_Comparison_2011}.
It can be distinguished between methods based on the acoustic analogy and those applying a wave-splitting. 
The first method appears to be more restrictive and therefore less interesting for real configurations.  
The wave-splitting methods rely on a modal decomposition of the unsteady field and exclusively apply to cylindrical duct sections. Thereby the unsteady field is fitted to some eigenmodes. 
These could be the true eigenmodes or an approximation of them--for instance the flow is assumed inviscid and swirl-free. 
When the duct geometry and the flow are kept constant, eigenmodes describe the
one set of perturbations whose patterns periodically repeat along the duct
~\cite{Weckmuller_Acoustic_2014}.
The radial shapes of these patterns and the corresponding axial wavenumbers are solely given by the background flow and the duct geometry. 
Wave-splitting methods can be divided into the L-R method and the triple plane pressure mode matching (TPP) method~\cite{Ovenden2004}. 
While the L-R-method matches all primitive variables on one axial plane, the TPP method uses the pressure on at least three planes.
Giacch\'e et al.~\cite{Giacche_Comparison_2011} showed that both methods perform equally well for rotor--stator-interaction noise.
The TPP method has the advantage that it can be most easily applied to
experimental data as those are obtained with pressure transducers (microphones).

As mentioned above, various eigensystems can be utilised for the wave splitting. 
The better the eigensystem matches the flow conditions, the more exact should be the calculation of the mode amplitudes. 
Clearly the limiting factor is the determination of the eigenmodes.

\begin{figure}[t]
	\centering
	\subfloat[\label{fig:TPP} ]
		{\includegraphics[height=0.34\textheight]{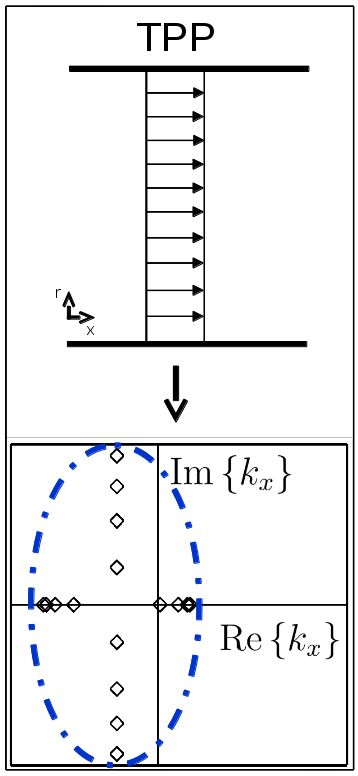}}\hfill
	\subfloat[\label{fig:TPPvilenski}]
		{\includegraphics[height=0.34\textheight]{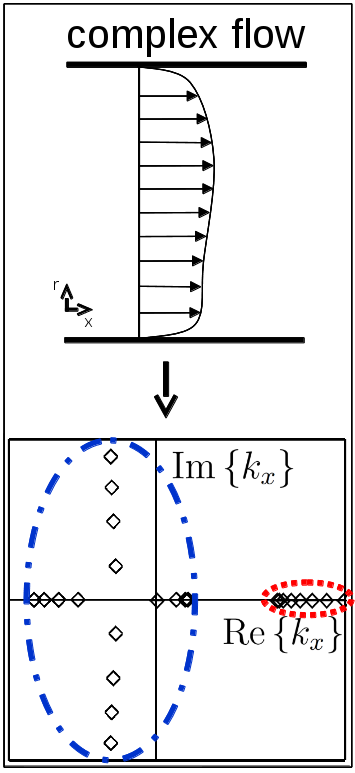}}\hfill
	\subfloat[\label{fig:XTPP}]
		{\includegraphics[height=0.34\textheight]{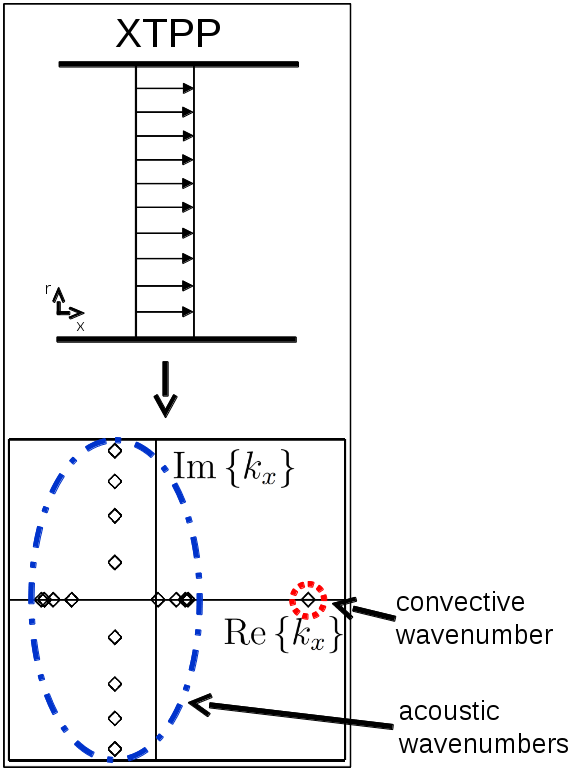}}
	\caption{\label{fig:overview}
	Radial mean flow profiles and corresponding axial wavenumber spectra; 
	({\color{blue}\bf - $\cdot$ -}) acoustic and ({\color{red}\bf - -}) convective wavenumbers. Note that in the XTPP method the convective wavenumber is added to the model.
	}
\end{figure}

The simplest available eigensystem is derived from the wave-equation for uniform mean flow and constant duct sections, see Fig.~\ref{fig:TPP}. 
In the last decades a rich body of literature was devoted to the formulation of the eigenmode analysis of the in-duct sound propagation on
sheared or swirling mean flows, starting 1958 with Pridmore-Brown~\cite{Pridmore-Brown_sound_1958}
up to the recent days, e.g.~\cite{Golubev1998}\nocite{Tam1998}\nocite{Moinier_Eigenmode_2005}\nocite{Boucheron_Analytical_2006}\nocite{Vilenski2007}--\cite{Chen_FourierBessel_2013}. 

Giacch\'e et al.~\cite{Giacche_Comparison_2011} showed that in real fan applications there are no significant differences found in the amplitudes of the blade passing frequency (BPF) determined by wave-splitting methods,
when applying either simple or complex eigenfunctions. This suggests that moderate flow non-uniformities do not significantly affect the predicted noise levels.

\begin{figure}[t]
\begin{centering}
\begin{tabular}{cccc}
\hskip -2mm\includegraphics[width=0.249\columnwidth]{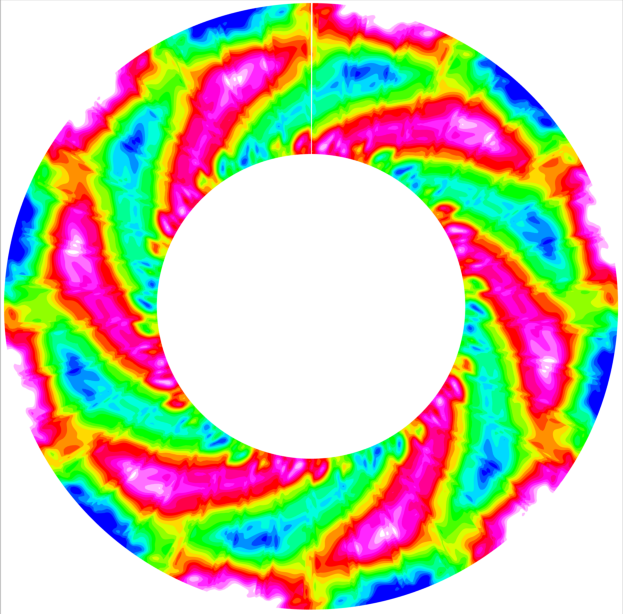} &
\hskip -4mm\includegraphics[width=0.249\columnwidth]{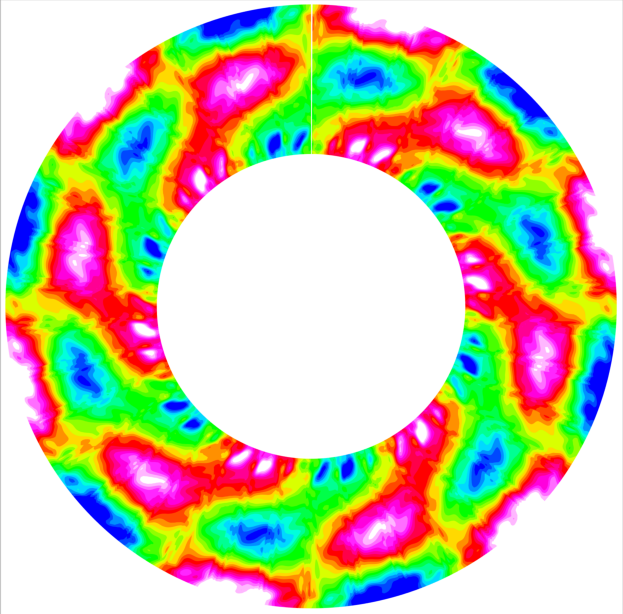}&
\hskip -4mm\includegraphics[width=0.249\columnwidth]{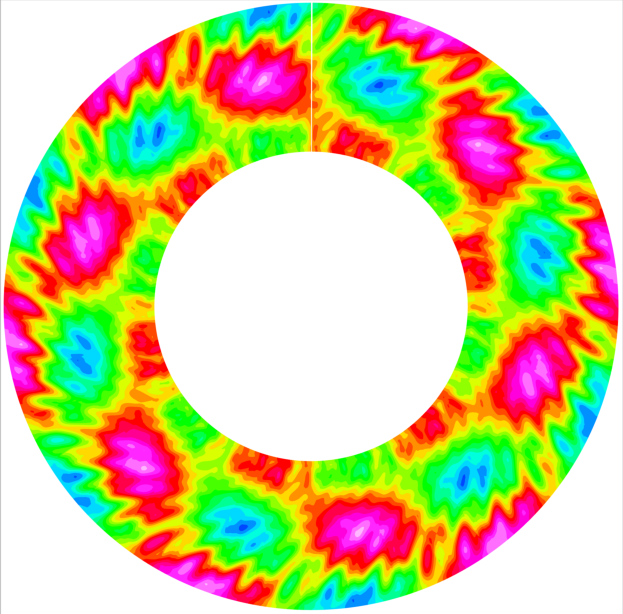}&
\hskip -4mm\includegraphics[width=0.249\columnwidth]{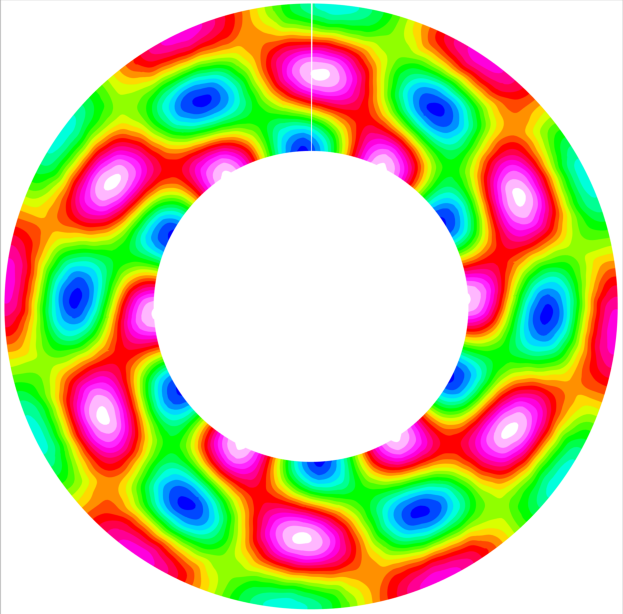} \\
 $x/c$=0.32 & $x/c$=0.5 & $x/c$=1 & $x/c$=2
\end{tabular}
\end{centering}
\caption{\label{fig:CFDslice1}
Example of pressure field at 2$\times$BPF issued from an unsteady RANS calculation; results downstream of the stator at 4 different axial positions normalised with the stator chord $c$ at midspan.}
\begin{centering}
\begin{tabular}{ccc}
\hskip -2mm\includegraphics[width=0.3\columnwidth]
{p_2_0261_color.png}&
\hskip -4mm\includegraphics[width=0.3\columnwidth]
{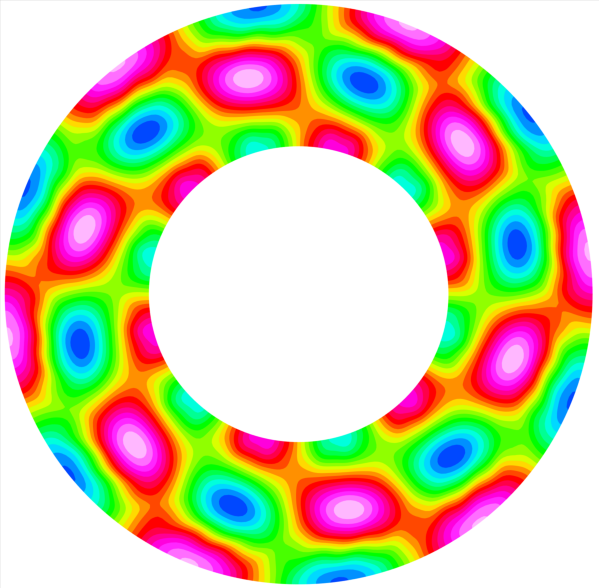} & 
\hskip -4mm\includegraphics[width=0.3\columnwidth]
{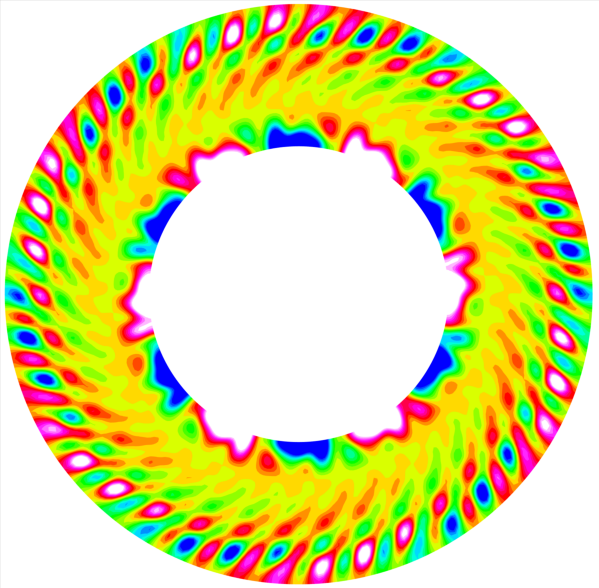}\\ 
raw CFD-pressure & acoustic part &	convective part\\
\end{tabular}
\caption{\label{fig:Recon}
(left) Raw CFD-pressure field and decomposition into (middle) acoustic and (right) convective parts ($x/c=1$). 
The scaling of the aerodynamic perturbations is changed to emphasise the structure.}
\end{centering}
\end{figure}

On the contrary the prediction of the sound power levels provided by the mode analysis reveals to be strongly dependent of the axial position of the analysis, we think because of the pseudo-sound contaminaton.
In Fig.~\ref{fig:CFDslice1} the pressure data at the first harmonic of the blade passing frequency from a URANS calculation are shown for different axial positions downstream of a stator. 
Details on the computation will be given in Section~\ref{sec:results}.
The perturbed flow field of a URANS calculation is modelled by non-linear viscous equations--full compressible Navier-Stokes equations with a turbulence model.
The Navier-Stokes equations describe the excitation, propagation, interaction 
 and destruction of vortical, entropic and acoustic perturbations~\cite{Chu1958}. 
Therefore the pressure field is composed of not only acoustic but also small scale perturbations, which are not acoustic in nature but travel downstream with convection speed.

In the literature the issue with pseudo-sound is handled by performing the mode analysis sufficiently far downstream of the stator trailing edge expecting the convective modes to vanish fast.
In some cases the user is constrained to perform several modal decompositions at growing distance to the sources, looking for the right position. 
This approach is time consuming and error prone. Futhermore in most application cases the space behind the stage is limited, not to think of a modal decomposition between two stages.

Our proposal is to introduce a new degree of freedom into the TPP method using a basis of functions for convective perturbations which helps then better filtering out pseudo-sound. 
Only one publication, by Vilenski~\cite{Vilenski2006}, is known to the authors where convective eigenmodes are taken into account in the pressure mode matching. These convective eigenmodes
results from the fact that the mean flow is no more uniform. This leads additionally to clustered convective wave numbers (Fig.~\ref{fig:TPPvilenski}). Vilenski showed that  
adding these convective components into the modal decomposition does not significantly improve the results of
the matching but instead can make it unstable as the eigenfunctions tend to be linearly dependent. 
For the case of swirling flow infinite families of such modes can
exist as shown by Golubev \& Atassi~\cite{Golubev_Sound_1996, Golubev1998} and Peake \& Parry~\cite{Peake_modern_2012}.
Thus it is most important to choose a good set of eigenmodes to keep the effort reasonable.

The extension proposed here is simple and robust. As illustrated in Fig.~\ref{fig:XTPP}, it retains the plug-flow assumption but adds a single convective wavenumber 
to contruct convective pseudo-modes. In our mind ``pseudo-modes'' are {\it not} eigenmodes of a certain set of partial differential equations but 
are issued from a model. They are similar to the acoustic modes and therefore can be used additionally in the pressure-mode matching method. 

The effect of this additional basis is the filtering of the convective components as showed in Fig.~\ref{fig:Recon}. 
The beneficial effect on acoustics is particularly obvious when comparing the acoustic part in Fig.~\ref{fig:Recon} to the raw CFD field in Fig.~\ref{fig:CFDslice1} measured at twice the distance to the stator
where convective perturbations have vanished. Both pressure patterns match very well.

We show in this paper that the new extension improves the results of the acoustic analysis in many ways:
\begin{itemize}
\item The amplitudes of the acoustic modes vary less in axial direction, improving interpretation and meaning.
\item The method enables the acoustic analysis in regions with significant convective disturbances.
\item The splitting of the convective from the acoustic fluctuating field is made possible.
\item The method gives identical results to the TPP method of Ovenden and Rienstra~\cite{Ovenden2004}, if no convective components are present.
\end{itemize}
Subsequently, we refer to this new method as the e{\bf X}tended-{\bf T}riple-{\bf P}lane-{\bf P}ressure-mode-matching (XTPP) method.

The paper is structured as follows. The XTPP method is described in Section~\ref{sec:method} and then applied to post-process a CFD simulation in Section~\ref{sec:results}. 
The physical interpretation of the convective pseudo-modes is discussed in Section~\ref{sec:discussion}.

\section{THEORY}
\label{sec:method}
The derivation of the e{\bf X}tended {\bf T}riple {\bf P}lane {\bf
P}ressure mode matching (XTPP) method is based on the
triple plane pressure mode matching (TPP) method proposed by Ovenden and Rienstra~\cite{Ovenden2004}. 
The technique is based on the modal decomposition at three adjacent axial planes $(x_0, x_1, x_2)$
and can handle with the case of ducts of slowly varying cross section. 
For means of clarity the extension of the propagation model to convective components is showed 
for a duct with constant cross-section only. 
The application to slowly-varying duct modes is straight forward.

\subsection{Acoustic modes for uniform mean flows}
\label{sec:moden}
The pressure field of the acoustic wave equation on uniform mean flows
in annular ducts has the following form: 
\begin{equation}
\label{eqn:modenTime}
p_a'(x,r,\theta,t)= \operatorname{Re} \left\{
\sum\limits_{m=-\infty}^{\infty} \sum\limits_{n= 0}^{\infty}
A^\pm_{mn}e^{i(k^\pm_{x,mn}x + m\theta - \omega t)}f_{mn}(r) 
\right\},	
\end{equation}
where $A^\pm_{mn}$ denotes the amplitudes of the upstream $(-)$ and downstream $(+)$ 
propagating modes, respectively. The wavenumber $k^\pm_{x,mn}$ is the axial
wavenumber and the function $f_{mn}(r)$ is the (radial) eigenfunction.
The numbers $m\in {\mathbb Z}_0$ and $n\in {\mathbb N}_0$ denote the azimuthal and 
the radial mode order, respectively. 
The normalised radial eigenfunctions $f_{mn}(r)$ consist of Bessel- and Neumann functions and read
\begin{equation}
\label{eqn:modenF}
f_{mn}(r) = \frac{1}{\sqrt{N_{mn}}}
\left(J_{m}\left(\sigma_{mn}\frac{r}{R}\right)
+Q_{mn}Y_{m}\left(\sigma_{mn}\frac{r}{R}\right)\right)\,,
\end{equation}
with $\sigma_{mn}$ and $Q_{mn}$ being defined by the boundary conditions 
at the inner and the outer duct radius respectively. 
The solution for hard walls is given in \ref{app:radEig}.
The axial wavenumbers $k_{x,mn}$ follow from the dispersion relation:
\begin{equation}
\label{eqn:moden_kx}
k^\pm_{x,mn} =\displaystyle \frac{k}{1-M_x^2}
\left ( -M_x\pm \sqrt{1-(1-M_x^2) \frac{\sigma_{mn}^2}{(kR)^2}}\right ).
\end{equation}
These can be simplified by introducing the cut-on factor
\begin{equation}
\label{eqn:moden_alpha}
 \alpha_{mn} = \sqrt{1-(1-M_x^2)\frac{\sigma_{mn}^2}{(kR)^2}}.
\end{equation}
Hence,
\begin{equation}
k^\pm_{x,mn} = \displaystyle \frac{k}{1-M_x^2}\left ( -M_x\pm \alpha_{mn} \right ).
\end{equation}
The sound power propagating in the duct is~\cite{Weckmuller2013,Moreau2011}:
\begin{equation}
\label{eqn:soundPower}
P^\pm_{mn} = \pi R^2 \Re{\frac{\alpha_{mn} (1-M_x^2)^2}{\bar \rho \bar c (1\mp M_x \alpha_{mn})^2 }}
|A^\pm_{mn}|^2 \,.
\end{equation}

\subsection{Construction of the convective model}
\label{sec:convModen}
Downstream of the acoustic source regions 
the non-stationary flow field is composed of acoustic pressure perturbations $p_a'$ as defined in Eq.~\eqref{eqn:modenTime} 
but also entropic and vortical perturbations called convective components and denoted $p_c'$. 
When only acouctic modes are used in the modal expansion as it is done by the TPP-method 
the modal amplitudes vary 
when the axial position of the three planes is varied. 
Thus the modal amplitudes and therefore the computed sound power which is propagated 
along the duct can strongly depend on the axial position 
where the matching is performed.

For simple problems at low Helmholtz number, where only plane waves are encountered, 
De Roeck~\cite{DeRoeck2006} suggested to add an aerodynamic model to the propagation model
used in the modal expansion. We generalise this concept  
to high Helmholtz numbers. 

The construction of the convective model is based
on the idea that pseudo-sound is convected by the mean flow. 
As illustrated in Fig.~\ref{fig:TPP}
the wavenumber spectrum of 
the linearized Euler equations with uniform mean flow does not contain such convective components. 
Thus it should be clear that the suggested convective model is not a mathematical solution of the wave equation. 
Its existence is physically motivated and turns out to be useful to minimise 
the errors resulting from the matching 
of the linear inviscid model to the non-linear viscous model of the CFD-domain.
Thus the convective wavenumber
\begin{equation}
\label{eqn:XTPP_kv}
k^c_x=\frac{k}{M_x}\;
\end{equation}
is introduced to model the propagation of the convective perturbations in the axial direction, 
where $k$ is the free-field wavenumber and $M_x$ is the axial Mach number 
of the uniform mean flow. 
The choice of the radial shape of the convective perturbations is free. We decided to use  
the same radial eigenfunctions $\eigf_{mn}(r)$ defined in Eq.~\eqref{eqn:modenF} as for the acoustics. Indeed these functions have this nice property of being orthogonal.
One could also imagine to use cosine functions.

The pressure field of the convective components $p_c'$ can then be written in analogy to Eq.~\eqref{eqn:modenTime} as:
\begin{equation}  
p_c'(x,r,\theta,t)= \operatorname{Re} \left\{
\sum\limits_{m=-\infty}^{\infty} \sum\limits_{n= 0}^{\infty}
\amp^c_{mn} e^{i(k^c_x x + m\theta - \omega t)}\eigf_{mn}(r) 
\right\}.
\end{equation}

\subsection{Application to the TPP-method}
\begin{figure}[hbt]
\begin{center}
\includegraphics[width=0.7\columnwidth]{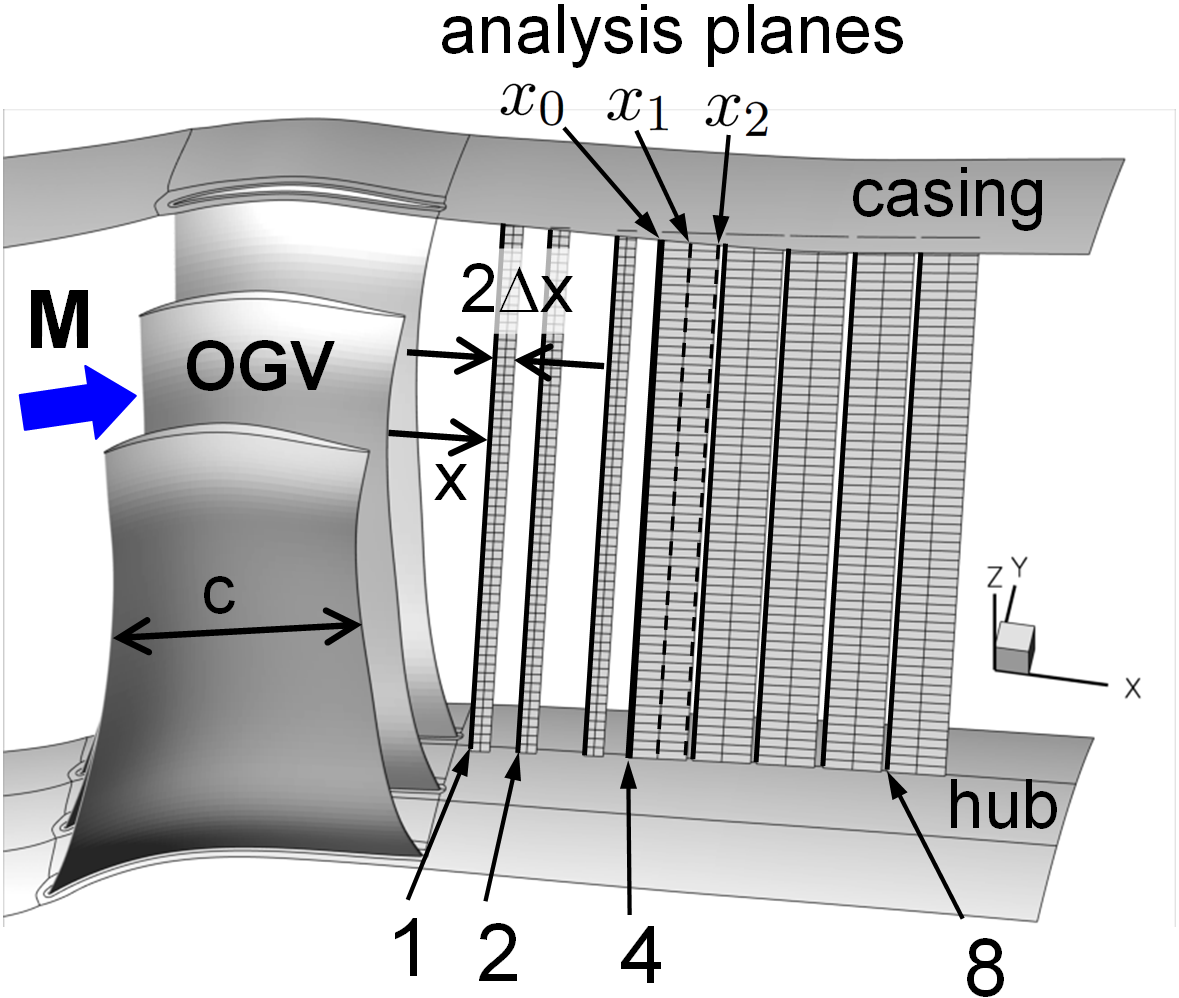}
\caption{\label{fig:setupSketch}
Placement of the three planes ($x_0,x_1,x_2$) 
at varying axial positions downstream of the OGV.}
\end{center}
\end{figure}

The extended triple-plane-pressure-mode-matching technique  
is derived similarly to Ovenden and Rienstra \cite{Ovenden2004} with adding a new basis for convective pressure perturbations.
The pressure field $p'$ is extracted from the CFD-domain (see Fig.~\ref{fig:setupSketch}) at three adjacent planes at axial positions $(x_0, x_1, x_2)$.
In a first step it is expanded in a Fourier series with respect to time 
and circumference:
\begin{equation}
p'(x,r,\theta,t) = \operatorname{Re} \left\{
\sum\limits_{m=-\infty}^{\infty} \sum\limits_{h= 0}^{\infty}
\hat p_{m\omega} (x,r)e^{i(m\theta - \omega t)}
\right\}.
\end{equation}
The complex pressure field at a single given frequency $\omega$ and azimuthal mode order $m$ 
is denoted by $\hat p_{m\omega}$.
Since the different $(m,\omega)$-components of the Fourier series are mutually independent, 
the derivation is done for one component of the Fourier series only. 
Thus $m$ and $\omega$ can be chosen arbitrary but are fixed 
and will not be noted explicitly in the following.
Consequently the pressure field at a given frequency $\omega$ and azimuthal order $m$ 
is written as a sum of upstream and downstream propagating acoustic modes 
and convected components:
\begin{subequations}
\label{eqn:_XTPP_fourierExpansion}
\begin{align}
\hat p_{m\omega} (x_0,r) = \hat p_0 (r)&=
\sum\limits_{n= 0}^{N}
\left(
\amp^\pm_{n}
+ \amp^c_n \right)
\eigf_{n}(r) \;, \\
\hat p_{m\omega} (x_1,r) = \hat p_1 (r)&=
\sum\limits_{n= 0}^{N}
\left(\amp^\pm_{n}e^{ik^\pm_{x,n} (x_1 - x_0)}
+ 
\amp^c_n e^{ik^c_x (x_1 - x_0)}\right)
\eigf_{n}(r) \;, \\
\hat p_{m\omega} (x_2,r) = \hat p_2 (r)&=
\sum\limits_{n= 0}^{N}
\left(\amp^\pm_{n}e^{ik^\pm_{x,n} (x_2 - x_0)}
+ 
\amp^c_n e^{ik^c_x (x_2 - x_0)}\right)
\eigf_{n}(r) \;.
\end{align}
\end{subequations}
The amplitudes $A^\pm_n$ of the acoustic modes of radial order $n$ and 
the amplitudes $A^c_n$ of the convective components are the unknowns. 
Compared to the TPP method the second term is new. 
The radial shape of the acoustic modes as well as the convective components 
are modelled by the same functions: 
the radial eigenfunctions $f_n$ of the wave operator as defined in Eq.~\eqref{eqn:modenF}.    
Similar to the TPP method for each radial mode order $n$
shape functions are defined to distinguish between upstream (-) and 
downstream (+) propagating acoustic modes in the projection procedure,
additionally we define shape function for the convective components (c):
\begin{subequations}
\label{eqn:XTPP_rescaling}
\begin{equation}
\xi^\dagger_{n}(r)= 
\left\{ \begin{array}{l @{\hspace{2cm}} l}
\displaystyle \eigf_{n}(r) 						&	(+) \\[6pt]
 \eigf_{n}(r)e^{ik^-_{x,n}(x_0-x_2)}  	&	(-) \\[6pt]
 \eigf_{n}(r)							 		&	(c)
 \end{array} \right.
\\[4pt]
\end{equation}
\begin{equation}
\zeta^\dagger_{n}(r)= 
\left\{ \begin{array}{l @{\hspace{2cm}} l}
\displaystyle \eigf_{n}(r)e^{ik^+_{x,n}(x_1-x_0)}	&	(+) \\[6pt]
 \eigf_{n}(r)e^{ik^-_{x,n}(x_1-x_2)}  			&	(-) \\[6pt]
 \eigf_{n}(r)e^{ik^c_{x}(x_1-x_0)} 	 			&	(c)
\end{array} \right.
\\[4pt]
\end{equation}
\begin{equation}
\chi^\dagger_{n}(r)= 
\left\{ \begin{array}{l @{\hspace{2cm}} l}
\displaystyle \eigf_{n}(r)e^{ik^+_{x,n}(x_2-x_0)}		&	(+)	\\[6pt]
 \eigf_{n}(r)  												&	(-) \\[6pt]
 \eigf_{n}(r)e^{ik^c_x(x_2-x_0)}						&	(c)	
\end{array} \right.
\end{equation}
\end{subequations}
Introducing these shape functions in the pressure series, see Eq.~\eqref{eqn:_XTPP_fourierExpansion},
new modal amplitudes arise. 
\begin{equation}
\label{eqn:XTPP_rescalingA}
B^\dagger_{n}= 
\left\{ \begin{array}{l @{\hspace{2cm}} l}
\displaystyle 
\amp^+_{n}							&	(+)	\\[6pt]
\amp^-_{n}e^{ik^-_{x,n}(x_2-x_0)}  	&	(-) \\[6pt]
\amp^c_{n}							&	(c)
\end{array} \right.
\\[4pt]
\end{equation}
To derive equations for the unknown modal amplitudes each equation of the pressure series 
at the three planes is multiplied by one of the adjoined shape functions and integrated over the radial coordinate:
\begin{subequations}
\label{eqn:XTPP_projection}
\begin{equation}
\sum\limits_{n= 0}^{N} B^{\ddagger}_{n}
\int\limits_{R_i}^{R_a}\xi^{\ddagger}_{n}(r)\,\xi^{\dagger^*}_{p}\!(r) \,r\!\dg\!r
=
\int\limits_{R_i}^{R_a}
\hat p_0 (r)\, \xi^{\dagger^*}_{p}\!(r) \,r\!\dg\!r\,,\\
\end{equation}
\begin{equation}
\sum\limits_{n= 0}^{N} B^\ddagger_{n}
\int\limits_{R_i}^{R_a}\zeta^\ddagger_{n}(r)\,\zeta^{\dagger^*}_{p}\!(r) \,r\!\dg\!r
=
\int\limits_{R_i}^{R_a}
\hat p_1 (r)\, \zeta^{\dagger^*}_{p}\!(r) \,r\!\dg\!r\,,\\
\end{equation}
\begin{equation}
\sum\limits_{n= 0}^{N} B^\ddagger_{n}
\int\limits_{R_i}^{R_a}\chi^\ddagger_{n}(r)\,\chi^{\dagger^*}_{p}\!(r) \,r\!\dg\!r
=
\int\limits_{R_i}^{R_a}
\hat p_2 (r)\, \chi^{\dagger^*}_{p}\!(r) \,r\!\dg\!r\,.
\end{equation}
\end{subequations}
where $p\in [0,N]$, with $N$ the maximal radial order, and  $(.)^*$ denotes
the complex conjugate.
A summation over $\ddagger \in [+,-,c]$ is implied.  
This can also be written in matrix notation:
\begin{equation}
\label{eqn:TPP_matrix}
{\bf M} {\bf a} = {\bf p}_0\, ,\quad 
{\bf N} {\bf a} = {\bf p}_1\, ,\quad 
{\bf Q} {\bf a} = {\bf p}_2\, .
\end{equation}
These are $3(N+1)$ equations at each axial position $(x_0,x_1,x_2)$. 
The three linear systems of equations show a block structure. 
The entries of a single block are defined as follows:  
\begin{subequations}
\label{eqn:XTPP_entries}
\begin{equation}
[{\bf M}]_{pn}=\!
\int\limits_{R_i}^{R_a}
\begin{pmatrix} 
\xi^+_{n}(r)\,\xi^{+^*}_{p}\!(r) &  \xi^-_{n}(r)\,\xi^{+^*}_{p}\!(r) &  \xi^c_{n}(r)\,\xi^{+^*}_{p}\!(r) \\[6pt]
\xi^+_{n}(r)\,\xi^{-^*}_{p}\!(r) &  \xi^-_{n}(r)\,\xi^{-^*}_{p}\!(r) &  \xi^c_{n}(r)\,\xi^{-^*}_{p}\!(r) \\[6pt] 
\xi^+_{n}(r)\,\xi^{c^*}_{p}\!(r) &  \xi^-_{n}(r)\,\xi^{c^*}_{p}\!(r) &  \xi^c_{n}(r)\,\xi^{c^*}_{p}\!(r) 
\end{pmatrix}\!
r\!\dg\!r
\,,\\[2pt]
\end{equation}
\begin{equation}
[{\bf N}]_{pn}=\!
\int\limits_{R_i}^{R_a}
\begin{pmatrix} 
\zeta^+_{n}(r)\,\zeta^{+^*}_{p}\!(r) & \zeta^-_{n}(r)\,\zeta^{+^*}_{p}\!(r) & \zeta^c_{n}(r)\,\zeta^{+^*}_{p}\!(r) \\[6pt]
\zeta^+_{n}(r)\,\zeta^{-^*}_{p}\!(r) & \zeta^-_{n}(r)\,\zeta^{-^*}_{p}\!(r) & \zeta^c_{n}(r)\,\zeta^{-^*}_{p}\!(r) \\[6pt]
\zeta^+_{n}(r)\,\zeta^{c^*}_{p}\!(r) & \zeta^-_{n}(r)\,\zeta^{c^*}_{p}\!(r) & \zeta^c_{n}(r)\,\zeta^{c^*}_{p}\!(r)
\end{pmatrix}\!
r\!\dg\!r
\,,\\[2pt]
\end{equation}
\begin{equation}
[{\bf Q}]_{pn}=\!
\int\limits_{R_i}^{R_a}
\begin{pmatrix} 
\chi^+_{n}(r)\,\chi^{+^*}_{p}\!(r) & \chi^-_{n}(r)\,\chi^{+^*}_{p}\!(r) & \chi^c_{n}(r)\,\chi^{+^*}_{p}\!(r) \\[6pt]
\chi^+_{n}(r)\,\chi^{-^*}_{p}\!(r) & \chi^-_{n}(r)\,\chi^{-^*}_{p}\!(r) & \chi^c_{n}(r)\,\chi^{-^*}_{p}\!(r) \\[6pt]
\chi^+_{n}(r)\,\chi^{c^*}_{p}\!(r) & \chi^-_{n}(r)\,\chi^{c^*}_{p}\!(r) & \chi^c_{n}(r)\,\chi^{c^*}_{p}\!(r)
\end{pmatrix}\!
r\!\dg\!r
\,,\\[2pt]
\end{equation}
\begin{equation}
[{\bf p}_0]_p=\!
\int\limits_{R_i}^{R_a}
\begin{pmatrix} 
\hat p_0 (r)\, \xi^{+^*}_{p}\!(r) \\[6pt]
\hat p_0 (r)\, \xi^{-^*}_{p}\!(r) \\[6pt]
\hat p_0 (r)\, \xi^{c^*}_{p}\!(r)
\end{pmatrix}\!
r\!\dg\!r\,,		\\[2pt]
\end{equation}
\begin{equation}
[{\bf p}_1]_p=\!
\int\limits_{R_i}^{R_a}
\begin{pmatrix} 
\hat p_1 (r) \zeta^{+^*}_{p}\!(r) \\[6pt]
\hat p_1 (r) \zeta^{-^*}_{p}\!(r) \\[6pt]
\hat p_1 (r) \zeta^{c^*}_{p}\!(r)
\end{pmatrix}\!
r\!\dg\!r\,,		\\[2pt]
\end{equation}
\begin{equation}
[{\bf p}_2]_p=\!
\int\limits_{R_i}^{R_a}
\begin{pmatrix} 
\hat p_2 (r) \chi^{+^*}_{p}\!(r) \\[6pt]
\hat p_2 (r) \chi^{-^*}_{p}\!(r) \\[6pt]
\hat p_2 (r) \chi^{c^*}_{p}\!(r)
\end{pmatrix}\!
r\!\dg\!r\,,\\[2pt]
\end{equation}
\begin{equation}
[{\bf a}]_n=
\begin{pmatrix} 
B^+_n \\[6pt]
B^-_n\\[6pt]
B^c_n
\end{pmatrix}.
\end{equation}
\end{subequations} 
To merge the three linear systems Eq.~\eqref{eqn:TPP_matrix} a cost function is defined:
\begin{equation}
\label{eqn:XTPP_costFunk}
f(\boldsymbol{a})=
||{\bf M} {\bf a} - {\bf p}_0||^2 +  
||{\bf N} {\bf a} - {\bf p}_1||^2 +  
||{\bf Q} {\bf a} - {\bf p}_2||^2\,.
\end{equation}
The vector of the unknown amplitudes ${\bf a}$ is determined by minimising this cost function.
The matrices ${\bf M}, {\bf N}, {\bf Q}$ are Hermitian due to their definition. 
Following Ovenden and Rienstra~\cite{Ovenden2004} this property is used 
to build one linear system of equations: 
\begin{equation}
\label{eqn:XTPP_LGS}
({\bf M}^2 + {\bf N}^2 + {\bf Q}^2)
{\bf a}
=
{\bf M} {\bf p}_0 +   
{\bf N} {\bf p}_1 + 
{\bf Q} {\bf p}_2\,.
\end{equation}
After introducing the extension of the propagation model by convective components 
we want to close this section with some remarks:
\begin{itemize}
  \item The aerodynamic model describes the convection of the perturbations 
    with the cross-sectional averaged mean flow, 
	but does not account either for the effects due to the wake expansion 
	nor for the viscous dissipation.
	The error due to these simplifications should be negligible if the evaluation planes 
	are closely spaced.
  \item While the axial wave numbers $k^\pm_{x,mn}$ of the acoustic modes are related 
	to their transverse wave numbers $k_{r,mn}$ by the dispersion relation, 
	the axial wave numbers of the convective perturbations $k^c_x$ are independent 
	of the azimuthal and radial mode orders. 
	Thus they depend only on the frequency and the Mach number. 
  \item While it is meaningful to speak of acoustic $(m,n)$-modes, 
	because these correspond to eigenfunctions of the wave equation,
	this is not the case for the convective components.
	Therefore, no physical interpretation should be be given to them: 
	the convective components just help to improve the mode matching and 
	enable to reconstruct the aerodynamic pressure field. 
	Nevertheless the sum can be interpreted as the whole convected pressure field.
  \item In case of orthogonality of the modal basis, 
    as it is the case for the Bessel- and Neumann eigenfunctions with hard-wall boundary conditions, 
    the matrices show a block-band structure.
    Therefore, only the amplitudes of the downstream and upstream propagating acoustic modes 
    and the convective components with the same radial order are related to each other.
\end{itemize}

\section{RESULTS}
\label{sec:results}
The pressure field resulting from a URANS computation of an UHBR fan stage is analysed 
using the TPP method and the present extension, XTPP. 
First some details on the CFD-simulation are given. 
Second the results of both methods are shown and compared to each other. 

\subsection{Numerical Simulation}
The unsteady flow field of the DLR UHBR fan~\cite{Kaplan2006} is calculated 
at approach condition (3187~rpm, mass flow 47.3~kg/s, PR=1.06) 
using the DLR CFD solver TRACE 
(Turbomachinery Research Aerodynamic Computational Environment~\cite{Nurnberger2001}) 
developed by the Department of Numerical Methods of the Institute of Propulsion Technology.
\begin{table}[htbp]
\begin{tabularx}{6.58cm}{
@{\arrayrulewidth1.5pt\vline\;}c
@{\;\arrayrulewidth1.5pt\vline\;}c|c|c|c
@{\;\arrayrulewidth1.5pt\vline}}
\noalign{\hrule height1.5pt} 
\multirow{2}{*}{Plane} & 
\multicolumn{4}{c@{\arrayrulewidth1.5pt\vline}}{Parameters}\\
\cline{2-5} 	& $x/c$ 	& $M_x$ 	& $R$	& $\Delta x/\lambda$ \\
	\noalign{\hrule height1.5pt}
	\hline		1	&  0.32			& 0.288 	& 0.420	 & 0.07\\
	\hline		2 	&  0.50			& 0.286 	& 0.422	& 0.07\\
	\hline		3 	&  0.75			& 0.283 	& 0.424	& 0.07\\
	\hline		4 	&  1.00			& 0.281 	& 0.427	& 0.21\\
	\hline		5 	&  1.25			& 0.278 	& 0.430	& 0.21\\
	\hline		6 	&  1.50			& 0.274 	& 0.432	& 0.21\\
	\hline		7 	&  1.75			& 0.270 	& 0.436	& 0.21\\
	\hline		8 	&  2.00			& 0.266 	& 0.439	& 0.21\\
	\noalign{\hrule height1.5pt}
\end{tabularx}
\centering
\caption{\label{tab:axPos}
Axial position of the evaluation plane $x/c$, axial Mach number $M_x$, outer duct radius $R$ and  
axial spacing between the 3 planes $\Delta x/\lambda$. 
The axial spacing is bigger for the last five positions since the underlying CFD-mesh coarsens 
with growing distance to the vane. 
}
\end{table}
The phase-lagged method~\cite{Gerolymos2001, Schnell2004} enables to reduce 
the computational domain to one passage for the rotor and one passage for the stator. 
Overall the multi-block structured grid consists of $3.3$ million nodes. 
The time integration is realised with 256 time steps per blade passing 
and 20 sub-iterations in the dual-time stepping algorithm using a Crank-Nicolson method. 
The fan is composed of 22 rotor blades and 38 stator vanes. 
The average axial Mach number $M_x$ varies between 0.288 right after the stator blades 
and 0.26 at the outlet plane (see Tab.~\ref{tab:axPos}). The fundamental blade passing frequency (BPF) is cut-off.  
More details on the URANS-simulation are given in Ref.~\cite{Weckmuller2009a}. 

In Figure~\ref{fig:CFDslice1} on page~\pageref{fig:CFDslice1} the pressure field
at the first harmonic of the blade passing frequency ($2\times$BPF), extracted at 4 axial positions, is shown. 
The distance $x$ denotes the spacing between the stator trailing edge at mid-span 
and the first of the three analysis planes at $x_0$. 
The chord length of the stator at mid-span is denoted by $c$. 
A dominant pattern with $|m|=6$ can be identified at each plane (see~Fig.~\ref{fig:CFDslice1}).
Close to the vanes additionally small-scale perturbations are observed. 
With increasing distance to the vanes, these small-scale perturbations tend to vanish. 
At $x/c=2$ the small-scale perturbations are not visible:  
the pressure field consists of a limited number of acoustic modes.
The disappearance of the small-scale perturbations is due to both physical effects 
modelled by the URANS equations (e.g. wake broadening) 
and the numerical dissipation resulting from the grid coarsening. 
With respect to the mode matching the reason of this is not important. 
We identify two domains of interest: a near-field domain close to the vanes, 
where the convective perturbations are of equal magnitude with the acoustic perturbations, and 
a far-field domain further downstream where the convective perturbations are negligible.

\subsection{Acoustic mode analysis} 
The acoustic mode analysis is performed at 8 different positions 
between $x/c=0.32$ and 2 (see Tab.~\ref{tab:axPos} and Fig.~\ref{fig:setupSketch} resp.). 
Each analysis is performed with three planes separated by a constant axial spacing $\Delta x$
specified in Tab.~\ref{tab:axPos}. 
Both the TPP and the XTPP method utilise the slowly-varying-duct modes~\cite{Rienstra1999} 
since the duct is slightly changing in radius. 
Any significant influence of the slowly-varying-duct modes on the radial mode analysis 
is excluded since the three planes are closely spaced.    
The first BPF harmonic ($2\times$BPF) is the strongest tone in the simulation. 
According to the rule of Tyler \& Sofrin~\cite{Tyler1962}, the rotor--stator interaction modes 
of azimuthal order $m=...,-44,-6,32,...$ 
should be present at $2\times$BPF\footnote{In this paper, positive mode orders
$m$ denote pressure patterns spinning in the positive $\theta$-direction of an orthogonal positive defined system of coordinates, 
whose $x$-axis is pointed in the direction of the mean flow.}.
The cut-on factor $\alpha_{mn}$, Eq.~\eqref{eqn:moden_alpha}, 
is used to separate cut-on and cut-off acoustic modes. 
The modes are cut-on as long as the cut-on factor is a real number. 
According to this, only the modes $(m,n)$=(-6,0), (-6,1), (-6,2), and (-6,3) 
are cut-on at $2\times$BPF. 
The modal decomposition presented hereafter was conducted with all modes 
with $-81 \leq m \leq 81$ and $0 \leq n \leq 10$, including many cut-off modes. 

First, to show the effect of the convective model extension, the results of 
the radial mode analysis are shown in terms of the modal amplitudes 
at a position $x_0$ one chord length downstream of the stator. 
Second, to demonstrate the benefit of the convective model, the results 
of the radial mode analysis are compared in terms of sound power in the duct (Eq.~\eqref{eqn:soundPower}) 
when the axial distance of 
the analysis planes to the stator is varied.   

\subsubsection{Effect of the convective model on the modal amplitudes}
\begin{figure}[p]
\centering
\subfloat[\label{fig:Aplus}Amplitudes of the downstream propagating acoustic waves.]{
\begin{tabularx}{\textwidth}{cXc}
\textbf{TPP} &&
\hspace{-1cm}\textbf{XTPP} \\\hline
\includegraphics[height=0.33\columnwidth]
{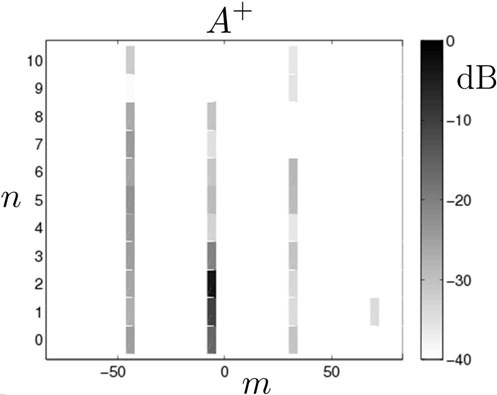} & &
\includegraphics[height=0.33\columnwidth]
{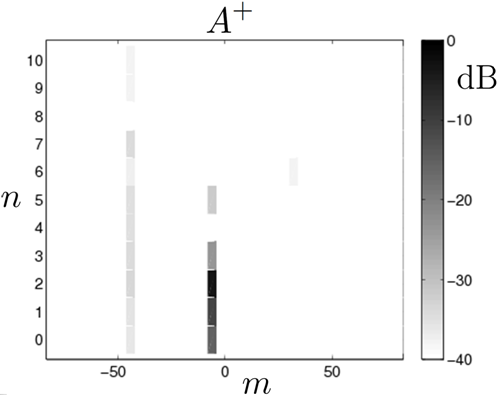}\\ 
\end{tabularx}}\\
\subfloat[\label{fig:Aminus}
Amplitudes of the upstream propagating acoustic waves.]{
\begin{tabularx}{\textwidth}{cXc}
\includegraphics[height=0.33\columnwidth]
{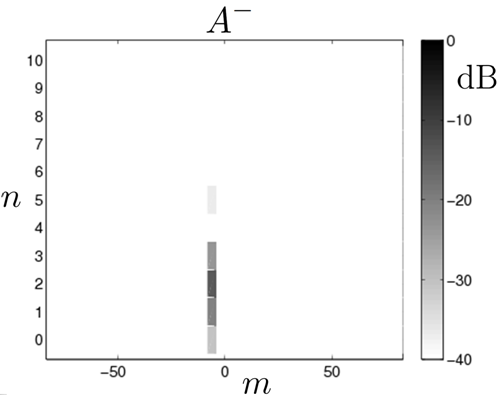} & &
\includegraphics[height=0.33\columnwidth]
{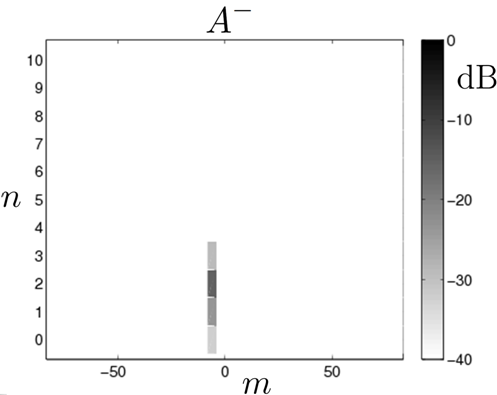}\\
\end{tabularx}}\\
\flushright
\subfloat[\label{fig:Aaero}
Amplitudes of the convective perturbations.]{
\begin{tabularx}{\textwidth}{cXc}
\includegraphics[height=0.33\columnwidth]
{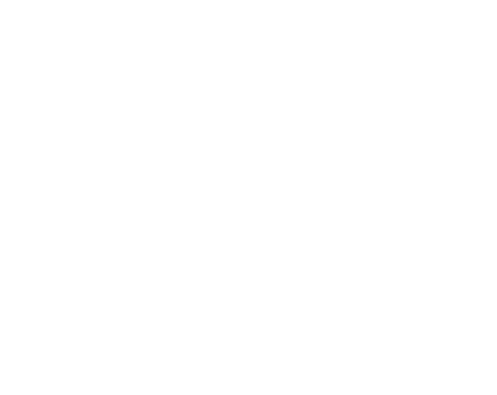} & &
\includegraphics[height=0.33\columnwidth]
{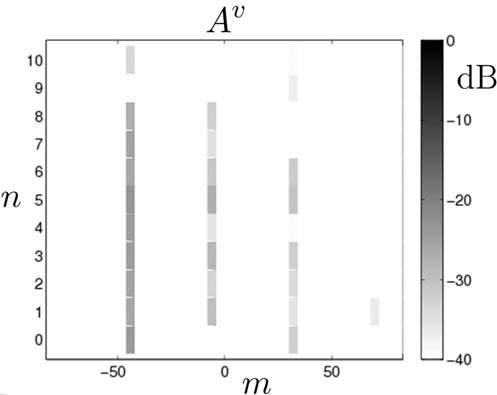}\\
\end{tabularx}}
\caption{\label{fig:A}Pressure amplitudes resulting from the modal decomposition
at $2\times$BPF and $x/c=1$.
The left column shows the results of the original TPP method proposed by Ovenden and Rienstra~\cite{Ovenden2004} 
while the right column shows the results of the extended method (XTPP).}
\end{figure}

Exemplarily the pressure amplitudes, resulting from the modal decomposition at $x/c=1$ are shown 
in Fig.~\ref{fig:Aplus}--\ref{fig:Aaero}. 
Significant levels are attributed only to the amplitudes $A^\pm_{mn}$ of the 
so-called Tyler and Sofrin modes.  
Comparing the amplitudes of the downstream propagating modes (Fig.~\ref{fig:Aplus}) 
resulting from the TPP or the XTPP method, 
it is obvious that the levels of the cut-off modes are significantly reduced 
while the levels of the cut-on modes remain roughly the same. 
The XTPP method now interpretes the modes with
$m=-44$, -6, 32 and 70 as convective perturbations (Fig.~\ref{fig:Aaero}).
The results of the XTPP method are meaningful since the distance to the stator is high and  
the amplitudes of the cut-off modes decreases exponentially.
The amplitudes of the cut-on modes, carrying the energy, remain nearly the same. The amount of 
significant cut-off modes is considerably decreased. 
The XTPP method filters out spurious noise, 
clarifying the interpretation of the acoustic results, 
even in regions without strong convective perturbations.

Based on the amplitudes of the acoustic and the convective model the pressure field 
belonging to both phenomena can be reconstructed separately. 
In Fig.~\ref{fig:Recon} on page~\pageref{fig:Recon} the raw CFD-pressure field
and the reconstructed acoustic and convective pressure fields are shown.
The most obvious result is that the reconstructed acoustic field at $x/c=1$ 
and the raw CFD data at $x/c=2$ match each other 
(compare Fig.~\ref{fig:CFDslice1} to Fig.~\ref{fig:Recon}). 
This highlights the capability of the XTPP method to remove the convective components 
from the raw CFD data. 
Furthermore, the convective part can easily be related to the rotor wakes 
scattered by the vanes.  
In the outer duct area between mid-span and casing 
the convective part shows a superposition of two structures with $|m|=44$ and $|m|=6$.    
At $2\times$BPF the wakes of the 22 rotor blades, rotating against the $\theta$-direction 
of a mathematically positive defined system of coordinates, are seen at $m=-44$.  
This $m=-44$ pattern is scattered into structures with $m=\ldots,-6,32,\ldots$ at the stator. 
This mode-scattering rule holds whatever the physical nature of the perturbations is, 
being acoustical, vortical or entropic. 
Thus the structure $|m|=6$ visible in the outer and inner duct area can be related to 
the blade tip and hub vortices of the rotor being scattered by the 38 stator vanes.

\subsubsection{Benefit of the convective model on the sound power levels}
\begin{figure}[!b]
\begin{center}
\includegraphics[width=0.85\columnwidth]{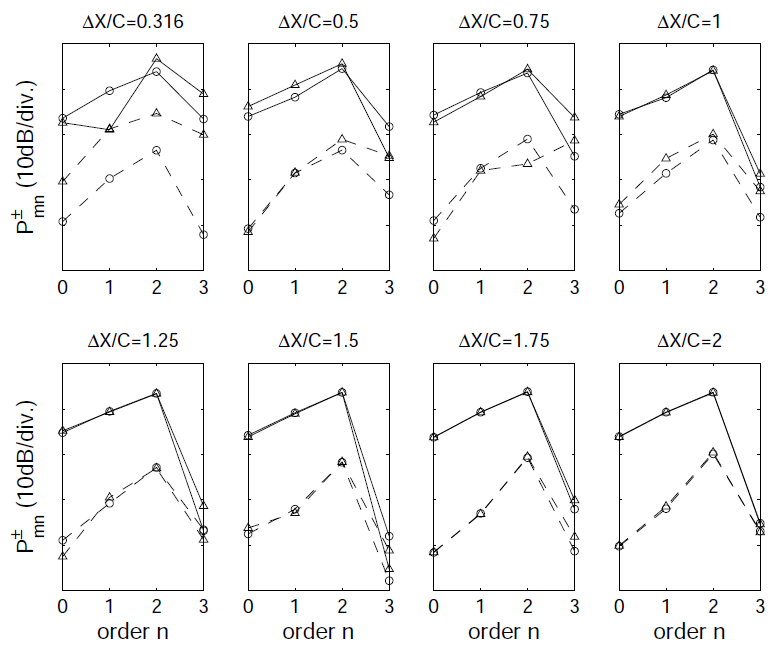}
\caption{\label{BPF2_3Planes_plot1}
Sound power amplitudes of the downstream (solid) and upstream (dashed) cut-on modes of same azimuthal order $m=-6$ as obtained at eight different positions downstream of the stator; 
($\vartriangle$) TPP, ($\circ$) XTPP. 
}
\end{center}
\end{figure}

The sound power carried by a $(m,n)$-mode along the duct can be expected to 
remain approximately constant when the mean flow 
and the duct contours vary little along the duct.  For the four cut-on modes at
$2\times$BPF this axial evolution is shown in Figure~\ref{BPF2_3Planes_plot1}. 
As shown previously, the mode analysis with the TPP and XTPP methods 
give equal results for the acoustic pressure amplitude 
far downstream where the convective perturbations are negigible. 
This confirms that the convective model does not alter the results. 

\begin{figure}[!ht]
\begin{center}
\includegraphics[width=0.7\columnwidth]{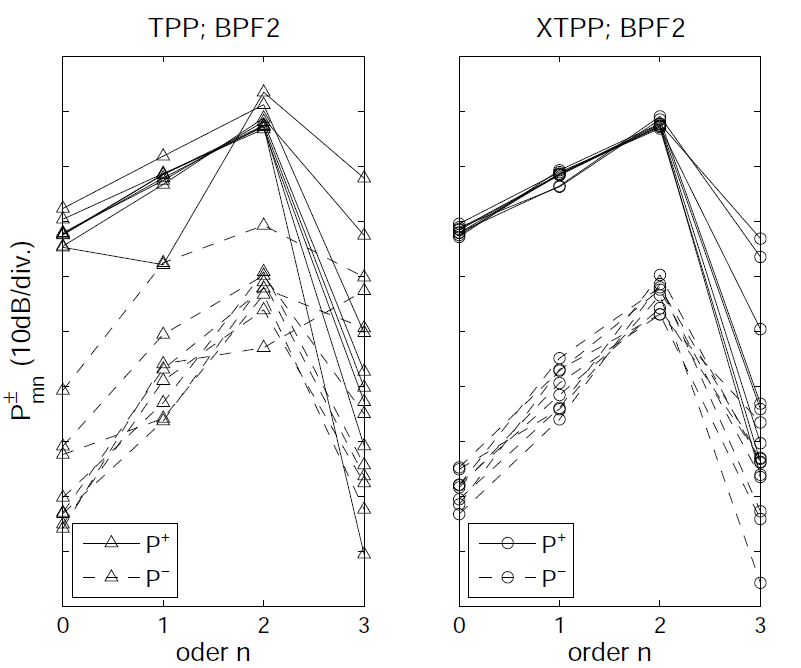}
\caption{\label{BPF2_3Planes_plot2}
Sound power amplitude of the downstream (solid) and upstream (dashed) modes of azimuthal order $m=-6$; 
(left) TPP, (right) XTPP. 
}
\end{center}
\end{figure}

A different situation occurs when performing the mode analysis 
close to the stator. The results of the TPP and XTPP methods strongly 
diverge. The results of Fig.~\ref{BPF2_3Planes_plot1} are represented in a different way in Fig.~\ref{BPF2_3Planes_plot2} 
where the sound power levels computed by one method at the different axial positions 
are superimposed in a single plot to emphasise axial variations in the results. 
The downstream propagating modes are generated by the stage. 
The amplitude of the radial components $n=$0, 1 and 2 vary within 10~dB when calculated 
with the TPP method and only within 2~dB when calculated with the XTPP method. 
The upstream propagating modes are due to numerical reflections 
at the outlet boundaries of the CFD-domain which are not perfectly nonreflecting. 
Whatever the method used is, the amplitude variations of 
the upstream propagating modes are larger than those of the downstream propagating modes. 
Notice that the amplitude variations of the upstream propagating modes 
are reduced with the XTPP method too. 
 
It can be observed that the amplitudes of the \mbox{(-6,3)-mode} decreases 
along the duct which is typical for cut-off modes. According to Eq.~\eqref{eqn:moden_alpha} this mode should be cut-on.  
However the formula is strictly valid for a uniform mean flow in a duct of constant cross section. 
It is expected that this mode is actually cut-off in the simulation as those hypothesese are violated. For sure using the plug flow assumption is a limitation of the current model.

\section{DISCUSSION}
\label{sec:discussion}

In order to distinguish the different aspects of the matching problem 
we regard it as a two step procedure. 
The first step is to define a propagation model and 
the second step is to match the coefficients of that model 
to the CFD perturbed pressure field. 

As stated by Ovenden and Rienstra~\cite{Ovenden2004} for the case of an orthogonal modal basis, 
the linear system--Eq.~\eqref{eqn:XTPP_LGS}--relates only the amplitudes of the downstream 
and upstream propagating acoustic waves with the same orders $(m,n)$. 
Since the convective model is based on the same radial and azimuthal eigenfunctions, 
the orthogonality relations for both the radial and circumferential directions 
are conserved. 
Thus the XTPP and the TPP method yield the same block-banded matrix structure.
This structure relates only the amplitudes of the upstream and downstream
propagating acoustic $(m,n)$-modes and additionally the
convected pseudo-modes with identical mode orders $(m,n)$ to the perturbation pressure. This is an important feature of the XTPP method.

When using the TPP method the convective components of the perturbed pressure field 
are incorporated in the right-hand side of the linear system--Eq.~\eqref{eqn:XTPP_LGS}--and must be balanced by the amplitudes of the up- and downstream propagating modes at 
fixed mode orders $(m,n)$. 
By introducing the convective model, additional degrees of freedom are offered to the matching process. 
Since these additional degrees of freedom have a physical meaning, 
being a rough model of the rotor wake perturbations, 
the matching is robust even in case of strong convective perturbations 
in the CFD data.    

The following simplifications have been made: 
\begin{itemize}
  \item The convective model describes an axial convection of perturbations without 
  	any radial variation of the mean flow (convection with the uniform mean flow).
  \item The acoustic model does not account for 
	neither the radial profile of the mean flow nor the presence of a swirl. 
  \item The convective model does not account for any dissipation of the perturbations 
    (wake mixing/broadening).
\end{itemize}

The first two assumptions can be relaxed if a sheared mean flow is included in the propagation model. 
When a numerical eigenmode analysis is performed on a sheared mean flow, 
the acoustic and convective eigenmodes deliver a more physical 
description of the propagation of small perturbations along the duct. 
When these eigenmodes are used as propagation model in the TPP method, 
it should not be necessary to extend the propagation model by convective components, 
since they naturally arise in the spectrum of the wave operator 
as described by Vilenski~\cite{Vilenski2006}. 
But when those are used in a mode matching procedure two drawbacks 
for solving the linear system may occur. 
First, the orthogonality of the modal basis is lost and thus the block-band
matrix structure too.
Second, the radial eigenfunctions of the convective modes tend to be linearly dependent. 
This poses a major issue on the uniqueness of the linear system, 
because the condition number of the system matrix can easily grow up, 
leading to an ill-posed system. 
Especially, when the cost function is transformed to a linear system using the normal equation. 
Indeed when the matrix is squared, its 
condition number is  squared too.  
Vilenski~\cite{Vilenski2006} stated 
that only if the number of convective modes incorporated in the mode analysis 
is restricted to a small number, improvements of the 
results could be achieved.

\section{CONCLUSION}
The triple plane pressure mode matching method introduced 
by Ovenden and Rienstra~\cite{Ovenden2004} 
has been extended by a convective model
and thus takes into account aerodynamic pressure fluctuations 
related to the rotor and stator wakes.
The original and the extended TPP method were applied to a URANS CFD-simulation 
of a fan stage.
In the regions where the convective perturbations of the raw CFD data are negligible, the
extended method provides the same results for the cut-on acoustic modes as the original method but
reduces the amplitude of the cut-off modes.
In regions, where the convective perturbations are dominant, 
the extended method gives the same (correct) cut-on acoustic results than further downstream, 
while the original TPP method shows strong discrepancies of the acoustic results. 
Consequently, the extension reduces the uncertainty of the computed sound power propagating in the duct 
and enables a more reliable interpretation of the acoustic results. 
The acoustic mode amplitudes provided by the extended method can be used to reconstruct 
the acoustic perturbations very close to the source region. 
This can be further propagated through the duct using a CAA solver. 
Thus it enables to reduce the size of the CFD domain and the computational costs.

A consequent extension of the propagation model with respect to 
vortical and swirling mean flows should further improve the quality 
of the results of the mode matching method.






\bibliographystyle{model1-num-names}
\bibliography{bib/biblio.bib}

\appendix
\section{Radial eigenfunctions of the acoustic wave equation}
\label{app:radEig}
In Section \ref{sec:method} we use the normalised radial eigenfunctions $f_{mn}(r)$ 
of the acoustic wave equation with uniform mean flow.  
They are a sum of Bessel- and Neumann functions and defined as:
\begin{equation}\tag{\ref{eqn:modenF}}
f_{mn}(r) = \frac{1}{\sqrt{N_{mn}}}
\left(J_{m}\left(\sigma_{mn}\frac{r}{R}\right)
+Q_{mn}Y_{m}\left(\sigma_{mn}\frac{r}{R}\right)\right)\,,
\end{equation}
with $\sigma_{mn}$ and $Q_{mn}$ being defined by the boundary conditions 
at the inner resp. the outer radius, here for the hard wall case:   
\begin{equation}
Q_{mn} = -\frac{J'_m(\sigma_{mn})}{Y'_m(\sigma_{mn})}\,.
\end{equation}
The normalisation factor $N_{mn}$ is defined such that:
\begin{equation}\label{eq:fmn_normalisation}
\frac{1}{2\pi R^2}\int_0^{2\pi}\int_{\eta R}^{R}|f_{mn}(r)|^2 r \mbox{d}r\mbox{d}\theta = 1\,.
\end{equation}
From the analytical solution of the integral it follows: 
\begin{equation}
 N_{mn} = \left\{\!\!\begin{array}{l l}
\displaystyle \frac{1}{2}[1-\eta^2] &\!\!\!\! ,\mbox{if}\,m=n=0 \\
\\[-6pt]
\displaystyle \frac{1}{2}\left [  \left (1-\frac{m^2}{\sigma_{mn^2}} \right )B_{mn}^2(R)-
\left (\eta^2-\frac{m^2}{\sigma_{mn^2}} \right ) B_{mn}^2(\eta R) \right ] & \!\!\!\!\mbox{otherwise}\,.\\
\end{array} \right.
\end{equation}
The radial eigenfunction is denoted by 
\begin{equation}
B_{mn}(r) = J_{m}\left(\sigma_{mn}\frac{r}{R}\right)
+Q_{mn}Y_{m}\left(\sigma_{mn}\frac{r}{R}\right)
\end{equation}
and the inner and outer radius of the duct are denoted by $\eta R$ and $R$ respectively. 
%







\end{document}